\documentstyle[12pt]{article}

\catcode `\@=11
\@addtoreset{equation}{section}
\def\theequation{\arabic{section}.\arabic{equation}}
\catcode `\@=12

\newcommand{\be}{\begin{equation}}
\newcommand{\en}{\end{equation}}
\newcommand{\bea}{\begin{eqnarray}}
\newcommand{\ena}{\end{eqnarray}}
\newcommand{\beano}{\begin{eqnarray*}}
\newcommand{\enano}{\end{eqnarray*}}
\newcommand{\bee}{\begin{enumerate}}
\newcommand{\ene}{\end{enumerate}}

\newcommand{\R}{R \!\!\!\! R}
\newcommand{\N}{N \!\!\!\!\! N}
\newcommand{\Z}{Z \!\!\!\!\! Z}

\newcommand{\I}{{\cal I}}

\newcommand{\Lc}{{\cal L}}

\newcommand{\Sc}{{\cal S}}
\newcommand{\Scmin}{{\em s}}

\begin{document}

\thispagestyle{empty}
 
\vspace*{1cm}

\begin{center}
{\Large \bf Multi-resolution analysis generated by a seed function}   \vspace{2cm}\\

{\large F. Bagarello}
%\footnote[1]{ Dipartimento di Matematica ed Applicazioni, 
%Fac. Ingegneria, Universit\`a di Palermo, I-90128  Palermo, Italy}  
\vspace{3mm}\\
  Dipartimento di Matematica ed Applicazioni, 
Fac. Ingegneria, Universit\`a di Palermo, \\I-90128  Palermo, Italy\\
e-mail: bagarell@unipa.it
\vspace{2mm}\\
\end{center}

\vspace*{2cm}

\begin{abstract}
\noindent 
In this paper we use the equivalence result originally proved by the author which relates a multi-resolution analysis (MRA) of $\Lc^2(\R)$  and an orthonormal set of single electron wave-functions in the lowest Landau level, to build up a procedure which produces, starting with a certain square-integrable function, a MRA of $\Lc^2(\R)$.     \end{abstract}

\vspace{2cm}

{\bf AMS Subject Classifications (200)}:  42A16, 42C40, 81V70

\vspace{.3cm}


\vspace{.3cm}

{\bf Running title}: Multi-resolution analysis from a seed function

\vfill

\newpage

% Section 1
\section{Introduction}

In a series of recent papers, \cite{bag1,bag2,bag3}, we have shown the existence of a relation between any MRA  of $\Lc^2(\R)$ and an orthonormal (o.n.) set of functions of $\Lc^2(\R^2)$ which (1) belong to the lowest Landau level (LLL), (2) are closed under the action of two commuting unitary translation operators, and 3) can be used to produce a normalized trial ground-state for the gas of $N$ electrons. This method has been used up to now to produce different trial ground states for the well known fractional quantum Hall effect (FQHE). In our original papers we were mainly interested in using known fact from MRA in order to get information about FQHE. However, already in \cite{bag1,bag2}, we have also discussed the possibility of reversing the construction, in order to get the coefficients of a MRA, in the sense of \cite{mal,heij},  simply starting from a given single electron o.n. basis closed under the action of two (magnetic) translation operators. To implement this proposal we only need such a set of wave-functions: then we immediately have the coefficients of the related MRA, \cite{bag1,bag2}. However, this approach is not really easy to be used, the reason being that there are not many examples of this kind of wave-functions in the LLL in the literature, \cite{bms,ferr}. 

 In this paper we consider a different possibility. We will show how a given function of $\Lc^2(\R)$ satisfying some extra condition, can be used to generate a set of coefficients related to a MRA of $\Lc^2(\R)$, \cite{mal,heij}.

The paper is organized as follows:

in the next section we quickly review the method proposed in \cite{bag1,bag2}, without insisting too much on its physical aspects. 

In Sections III and IV we show how to use a seed function in order to construct a set of coefficients giving rise to a MRA.

In Section V we discuss some examples, and we discuss our conclusions in Section VI.

In the Appendix we prove some easy results on the convolution of sequences which are used in the main body of the paper, results which we were not able to find in the existing literature.

% Section 2
\section{The method}

We begin this section with the following remark: in  \cite{bag2,bag3} the method originally introduced in \cite{bag1} has been generalized. This generalization, which is crucial for concrete applications in the analysis of the FQHE, is only an unnecessary complication here and, for this reason,  will not be used.

The many-body model of the FQHE  consists simply in a two-dimensional electron gas, 2DEG, (that is a gas of electrons constrained in a two-dimensional layer) in a positive uniform background and subjected to an uniform magnetic field along $z$, whose hamiltonian (for $N$ electrons) is, \cite{bag1},
\begin{equation}
H^{(N)}=H^{(N)}_0+\lambda(H^{(N)}_C+H^{(N)}_B),\label{31}
\end{equation}
where $H^{(N)}_0$ is the sum of $N$ contributions:
\begin{equation}
H^{(N)}_0=\sum^N_{i=1}H_0(i).\label{32}
\end{equation}
Here $H_0(i)$ describes the minimal coupling of the $i-$th
electron with the magnetic field:
\begin{equation}
H_0={1\over 2}\,\left(\underline p+\underline A(r)\right)^2={1\over 2}\,\left(p_x-{y\over 2}\right)^2+{1\over 2}\,\left(p_y+{x\over 2}\right)^2. \label{33}
\end{equation}
$H^{(N)}_C$ is the canonical Coulomb interaction between charged
particles,
$
H^{(N)}_C={1\over2}\,\sum^N_{i\not=j}{1 \over|\underline r_i-
\underline r_j|},$
and $H^{(N)}_B$ is the interaction of the charges with the
background,  \cite{bms}. 

We now consider $\lambda(H^{(N)}_C+H^{(N)}_B)$ as a perturbation of the free hamiltonian $H^{(N)}_0$, and we  look for eigenstates of $H^{(N)}_0$ in the form of Slater determinants built up with single electron wave functions.  The easiest way to approach this problem consists in
introducing the  new variables
  \be
\label{35}
  P'= p_x-y/2, \hspace{5mm}     Q'= p_y+x/2.
  \en
In terms of $P'$ and $Q'$ the single electron hamiltonian, $H_0$, can be written as 
 \be
\label{36}
  H_{0}=\frac{1}{2}(Q'^2 + P'^2).
  \en
The transformation (\ref{35}) can be seen as a part of a canonical map from $(x,y,p_x,p_y)$ into $(Q,P,Q',P')$ where

   \be
\label{37}
   P= p_y-x/2, \hspace{5mm}    
  Q= p_x+y/2.
   \en
  These operators  satisfy the following commutation relations:
  \be 
\label{38}
 [Q,P] = [Q',P']=i, \quad  [Q,P']=[Q',P]=[Q,Q']=[P,P']=0.
  \en
   It is shown in \cite{daza,mo} that  a wave function in the $(x,y)$-space
is
   related to its  $PP'$-expression by the formula
  \be
\label{39}
  \Psi(x,y)=\frac{e^{ixy/2}}{2\pi}\int_{-\infty}^{\infty}\,
  \int_{-\infty}^{\infty}e^{i(xP'+yP+PP')}\Psi(P,P')\,dP dP',
  \en
which can be easily inverted:
  \be
\label{325}
  \Psi(P,P')=\frac{e^{-iPP'}}{2\pi}\int_{-\infty}^{\infty}\,
  \int_{-\infty}^{\infty}e^{-i(xP'+yP+xy/2)}\Psi(x,y)\,dx dy.
  \en
  The usefulness of the $PP'$-representation stems from the expression
(\ref{36}) 
  of $H_0$. Indeed, in this representation, the single electron Schr\"{o}dinger equation
admits 
  eigenvectors $\Psi(P,P')$ of $H_0$ of the form $\Psi(P,P')=f(P')h(P)$.
Thus
   the ground state of (\ref{36}) must have the form
 $f_0(P')h(P)$,  where
   \be     \label{310}
  f_0(P')= \pi^{-1/4} e^{-P'^2/2},
  \en
  while the function $h(P)$ is arbitrary, which manifests the degeneracy of
the LLL, and should be fixed by the interaction. With $f_0$ as above formula (\ref{39}) becomes
\be           
 \label{311}
  \psi(x,y) = \frac{e^{ixy/2}}{\sqrt{2}\pi^{3/4}}
\int_{-\infty}^{\infty}\,e^{iyP}e^{-(x+P)^2/2}h(P)\,dP, 
 \en
while, using (\ref{325}), $h(P)$ can be written in terms of $\psi(x,y)$ as
\be
h(P)=\frac{e^{-iPP'+P'^2/2}}{2\pi^{3/4}}\int_{-\infty}^{\infty}\,
  \int_{-\infty}^{\infty}e^{-i(xP'+yP+xy/2)}\Psi(x,y)\,dx dy
\label{326}
\en

Let us now define the so-called magnetic translation operators $T(\vec{a_i})$ for a square lattice with basis $\vec{a_1}=a(1,0)$, $\vec{a_2}=a(0,1)$, $a^2=2\pi$, \cite{bag1},  by
   \be
  T_1:=T(\vec{a_1})=e^{iaQ}, \quad T_2:=T(\vec{a_2})=e^{iaP}.
\label{312}
  \en
We see  that, due to (\ref{38}) and to the condition on the cell of the lattice, $a^2=2\pi$,
\be
[T(\vec{a_1}),T(\vec{a_2})]=[T(\vec{a_1}), H_0]=[T(\vec{a_2}),H_0]=0.
\label{314}
\en

The action of the $T$'s on a generic function $f(x,y)\in \Lc^2(\R^2)$ is the following
\be
f_{m,n}(x,y):=T_1^mT_2^n f(x,y)=(-1)^{mn}e^{i\frac{a}{2}(my-nx)}f(x+ma,y+na).
\label{316}
\en
This formula shows that, if for instance $f(x,y)$ is localized around the origin, then 
$f_{m,n}(x,y)$ is localized around the  site $a(-m,-n)$ of the square lattice.

Now we have all the ingredients to construct the ground state of $H^{(N)}_0$ mimiking the classical procedure. We simply start from the single electron ground state of $H_0$ given in (\ref{311}), $\psi(x,y)$. Then we construct a set of copies  $\psi_{m,n}(x,y)$ of $\psi$ as in (\ref{316}), with $m,n\in\Z$. All these functions still belong to the lowest Landau level for any choice of the function $h(P)$ due to (\ref{314}). $N$ of these wave functions $\psi_{m,n}(x,y)$ are finally used to construct a Slater determinant for the finite system:
\be
\psi^{(N)}(\underline r_1,\underline r_2,...,\underline r_N)= \!\! \frac{1}{\sqrt{N!}}\left|
\begin{array}{ccccccc}
\psi_{m_1,n_1}(\underline r_1) & \psi_{m_1,n_1}(\underline r_2) & . & . & . & . & \psi_{m_1,n_1}(\underline r_N)  \\ 
\psi_{m_2,n_2}(\underline r_1) & \psi_{m_2,n_2}(\underline r_2) & . & . & . & . & \psi_{m_2,n_2}(\underline r_N)  \\ 
. & . & . & . & . & . & .   \\ 
. & . & . & . & . & . & .   \\
. & . & . & . & . & . & .   \\
\psi_{m_N,n_N}(\underline r_1) & \psi_{m_N,n_N}(\underline r_2) & . & . & . & . & \psi_{m_N,n_N}(\underline r_N)  \\ 
\end{array}
\right|
\label{317}
\en

It is known, \cite{bms}, that in order to have $<\psi^{(N)},\psi^{(N)}>=1$ for all $N$ we need to have \be
<\psi_{m_i,n_i}\psi_{m_j,n_j}>=\delta_{m_i,m_j}\delta_{n_i,n_j}.
\label{317bis}
\en

Let $\psi(x,y)$ be as in (\ref{311}) and 
$\psi_{m,n}(x,y)=T_1^mT_2^n \psi(x,y)=(-1)^{mn}e^{i\frac{a}{2}(my-nx)}\psi(x+ma,y+na).
$
After few computations and using again  condition $a^2=2\pi$, we get
\be
\psi_{m,n}(x,y)=\frac{e^{i\frac{xy}{2}+iamy}}{\sqrt{2}\pi^{3/4}}\int_{-\infty}^\infty dP e^{i(y+na)P-(x+ma+P)^2/2}h(P).
\label{318}
\en
We have discussed in \cite{bag1} conditions on $h(P)$ such that equality (\ref{317bis}), or its equivalent form 
\be
\tilde S_{m,n}:=<\psi_{0,0},\psi_{m,n}>=\delta_{m,0}\delta_{n,0},\quad\forall m,n\in\Z,
\label{319}
\en
are satisfied. With the above definitions we find
\be
\tilde S_{l_1,l_2}=\int_{-\infty}^\infty dp e^{-il_2ap}\overline{h(p-l_1a)}h(p),
\label{320}
\en
which restates the problem of the orthonormality of the wave functions in terms of $h(P)$. In particular we see that, for $m=n=0$, this equation implies that $\psi_{00}$ in normalized in $\Lc^2(\R^2)$ if and only if $h(P)$ is normalized in $\Lc^2(\R)$. This reflects the unitarity of the transformation (\ref{39}), which, more in general, implies that any o.n. set in $\Lc^2(\R)$ is mapped into an o.n. set in $\Lc^2(\R^2)$.

In the construction above we are considering a square lattice in which all the lattice sites are occupied by an electron. We say that the {\em filling factor} $\nu$ is equal to 1. We have seen in \cite{bag1} that, in order to construct an o.n. set of functions in the LLL corresponding to a filling $\nu=\frac{1}{2}$ (only half of the lattice sites are occupied), we have to replace (\ref{319}) and (\ref{320}) with the following slightly weaker condition,
\be
S_{l_1,l_2}=\tilde S_{l_1,2l_2}=\int_{-\infty}^\infty dp e^{-2il_2ap}\overline{h(p-l_1a)}h(p)=\int_{-\infty}^\infty dp e^{il_1ap}\overline{\hat h(p-2l_2a)}\hat h(p)=\delta_{l_1,0}\delta_{l_2,0},
\label{321}
\en
for all $l_1,l_2\in\Z$, where $\hat h(p)=\frac{1}{\sqrt{2\pi}}\int_{\R}e^{-ipx}h(x)dx$ is the Fourier transform of $h(x)$. If $h(x)$ satisfies (\ref{321}), then, defining
\be
h_n=\frac{1}{\sqrt{a}}\int_{-\infty}^\infty dp e^{-inxa}h(x),
\label{322}
\en
it is easily checked that 
\be
\sum_{n\in\Z}h_n \overline{h_{n+2l}}=\delta_{l,0}.
\label{323}
\en
The proof of this claim, contained in \cite{bag1}, is based on condition (\ref{321}) and on the use of the Poisson summation formula (PSF) which we write here as 
\be
\sum_{n\in\Z}e^{inxc}=\frac{2\pi}{|c|}\sum_{n\in\Z}\delta(x-n\frac{2\pi}{c}).
\label{324}
\en
It is well known that the PSF does not always hold, and conditions for its validity are given in several papers and books, see \cite{chen} p.298  and references therein, for instance. In this paper we will always assume its validity, and from time to time we will check it explicitly. 

Equation (\ref{323}) shows how a function $h(x)$, satisfying the orthonormality condition (ONC) (\ref{321}) can be used to generate, via (\ref{322}), a set of coefficients which are related to a MRA, \cite{mal,heij,dau}. This procedure can be  extended in many ways which are not relevant here, \cite{bag1,bag2,bag3}, and therefore will not be considered in this paper. In \ref{bag1,bag2} is also discussed in some details the role of the Zak transform in our procedure, while a detailed summary of our results can be found in \cite{antbag}.

Several problems arise at this point:

1) is there any simple way to construct functions $h(x)$ which solves the ONC (\ref{321})? Of course, any o.n. basis $\Psi_{n,m}(x,y)$ arising in the analysis of the FQHE could be used to construct such a $h(x)$, but the literature is rather poor of these examples, \cite{bms,ferr}.

2) equation (\ref{323}) is not the only condition which should be satisfied by a set of complex numbers in order  to get a MRA of  $\Lc^2(\R)$, see \cite{mal,heij,dau} and the definition below. What can be said about the other conditions?

We will consider the first point above in the next section. Point 2) will be analyzed in Section IV.

\vspace{3mm}

We end this section by the following 

\noindent
\underline{\bf Definition:--} We call {\em relevant} any sequence $h=\{h_n, n\in\Z\}$ which satisfies the following properties:

\begin{itemize}
\item[(r1)]\hspace{2cm}
$\sum_{n\in\Z}h_n \overline{h_{n+2l}}=\delta_{l,0}$;
\item[(r2)]\hspace{2cm}
$h_n=O(\frac{1}{1+|n|^2}), \quad n\gg 1;$
\item[(r3)]\hspace{2cm}
$\sum_{n\in\Z}h_n =\sqrt{2};$
\item[(r4)]\hspace{2cm}
$H(\omega)=\frac{1}{\sqrt{2}}\sum_{n\in\Z}h_ne^{-i\omega n}\neq 0 \quad\quad \forall \omega\in[-\frac{\pi}{2},\frac{\pi}{2}].$
\end{itemize}

The role of relevant sequences in connection with MRA is explained in \cite{mal,heij,dau}, for instance, and will not be discussed here.

% Section 3
\section{The seed function, part one}

In this section we will show how to find, under very general assumptions, sequences satisfying condition (r1) above by making use of the approach outlined in the previous section. In particular we will show how, starting with a given {\em seed function} $h\in \Lc^2(\R)$, we can obtain another function $H$ satisfying the ONC (\ref{321}) and, as a consequence, a set of coefficients defined as in (\ref{322}) which satisfies condition (r1). As it will appear evident, a crucial role is played by formulas (\ref{39}) and (\ref{326}).

Let $h(P)$ be a generic square integrable function. Using formula (\ref{39}) we get a function $\Psi_h(x,y)=\frac{e^{ixy/2}}{\sqrt{2}\pi^{3/4}}
\int_{-\infty}^{\infty}\,e^{iyP}e^{-(x+P)^2/2}h(P)\,dP$ which belongs to the LLL independently of the choice of $h(P)$. Using $T_1$ and $T_2$ we define other functions, still belonging to the LLL, as in (\ref{316}):
\be
\Phi_{h,\underline l}(\underline r)=T_1^{l_1}T_2^{2l_2}\Psi_h(x,y)=e^{-i/2(\tilde X_{\underline l}y-\tilde Y_{\underline l}x)}
\Psi_h(\underline r-\underline{\tilde R}_{\underline l}),
\label{41}
\en
where we use the  notation $\underline l=(l_1,l_2)$, and we have defined $\underline{\tilde R}_{\underline l}=(\tilde X_{\underline l},\tilde Y_{\underline l})=-a(l_1,2l_2)$. Notice that, since we are considering even powers of $T_2$, we  obtain  a set of normalized wave functions of the LLL corresponding to a filling $\nu=\frac{1}{2}$ which are mutually orthogonal whenever  the seed function $h(P)$ satisfies the ONC (\ref{321}), \cite{bag1}, and, via (\ref{322}), also a set of coefficients $h_n$ satisfying (r1) above. However, in general, $h(P)$ does not satisfy (\ref{321}). We want to show here the way in which a function $H(P)$ satisfying the ONC can be obtained starting from this original $h$. The function $H(P)$ will be used to define some coefficients as shown in (\ref{322}). 

First of all we use the set $\I_\Phi=\{\Phi_{h,\underline l}, \underline l\in\Z^2\}$ to construct another set of functions, still belonging to the LLL, by considering the following superposition:
\be
\chi_{\underline n}(\underline r)=\sum_{\underline l\in\Z^2}f_{\underline l}\Phi_{h,{\underline l+\underline n}}(\underline r),
\label{42}
\en
where $\underline n=(n_1,n_2)$. The set $\I_\chi=\{\chi_{\underline n}, \underline n\in\Z^2\}$ shares with $\I_\Phi$ the property of being closed under the action of the magnetic translations:
\be
\chi_{\underline n}(\underline r)=T_1^{n_1}T_2^{n_2}\chi_{\underline 0}(\underline r).
\label{43}
\en
For this reason we can consider $\chi_{\underline 0}(\underline r)$ as a function in the LLL obtained from a $H(P)$, different from the seed function $h$, via the same transformation (\ref{39}), $\chi_{\underline 0}(\underline r)=\phi_{H}(\underline r)$, so that $H(P)$ can be obtained from $\chi_{\underline 0}(\underline r)$ by considering the inverse transformation (\ref{326}). The coefficients $f_{\underline l}$ will now be fixed by requiring that the set $\I_\chi$ is made of o.n. functions:
\be
<\chi_{\underline n},\chi_{\underline 0}>=\delta_{{\underline n},{\underline 0}}=\delta_{n_1,0}\delta_{n_2,0},
\label{44}
\en
for all integers $n_1$ and $n_2$. Using  (\ref{42}) and the following  equality,
\be
S^{(h)}_{{\underline l}+{\underline n}}=<\Phi_{h,{\underline l}+{\underline n}},\Phi_{h,\underline 0}>=<\Phi_{h,\underline l},\Phi_{h,-\underline n}>,
\label{45}
\en
which follows from the unitarity of $T_i$ and from (\ref{43}), the orthonormality constraint (\ref{44}) becames
\be
\sum_{\underline l,\underline s\in\Z^2}\overline{f_{\underline l}}f_{\underline s}S^{(h)}_{{\underline l}+{\underline n}-{\underline s}}=\delta_{\underline n,\underline 0}.
\label{46}
\en
Incidentally we recall that $S^{(h)}_{\underline l}$ can be rewritten in terms of the seed function as in (\ref{321}). We use here $S^{(h)}_{\underline l}$ instead of the simplest $S_{\underline l}$ to emphasize the role of the seed function $h$.  Introducing the following functions:
\be
F(\underline p)=\sum_{n\in\Z^2}f_{\underline n}e^{i\underline p\cdot \underline n}, \quad S^{(h)}(\underline p)=\sum_{n\in\Z^2}S^{(h)}_{\underline n}e^{i\underline p\cdot \underline n},
\label{47}
\en
equation (\ref{46}) can be rewritten as $|F(\underline p)|^2S^{(h)}(\underline p)=1$, whose solution is:
\be
F(\underline p)=\frac{e^{i\varphi(\underline p)}}{\sqrt{S^{(h)}(\underline p)}},
\label{48}
\en
$\varphi(\underline p)$ being a generic real function. To simplify the treatment, we will put $\varphi(\underline p)=0$ from now on. We will comment on this choice at the end of Section V. Notice that since the coefficients $S^{(h)}_{\underline n}$ satisfy the relation $S^{(h)}_{\underline n}=\overline{S^{(h)}_{-\underline n}}$, then $S^{(h)}(\underline p)$ is a real function, which is surely non negative. In order to avoid problems with possible divergences arising when $S^{(h)}(\underline p)=0$, we will try to consider in the following only those seed functions for which  $S^{(h)}(\underline p)$ is strictly positive.

Once the function $F(\underline p)$ is known, obtaining the coefficients $f_{\underline s}$ is quite straightforward:
\be
f_{\underline s}=\frac{1}{(2\pi)^2}\int_0^{2\pi}\int_0^{2\pi}d^2\underline p\frac{e^{-i\underline p\cdot \underline s}}{\sqrt{S^{(h)}(\underline p)}}.
\label{49}
\en
It is not difficult to check explicitly this result: if we use (\ref{49}) in the expansion (\ref{42}), we recover $<\chi_{\underline n},\chi_{\underline 0}>=\delta_{\underline n,\underline 0}$, as expected. In the proof of this statement the PSF has to be used.

 The coefficients $f_{\underline s}$ and equation (\ref{42}) produce a function $\chi_{\underline 0}(\underline r)$ which, together with its magnetic translated $\chi_{\underline n}=T_1^{n_1}T_2^{2n_2}\chi_{\underline 0}$, gives rise to an o.n. set in the LLL, for $\nu=\frac{1}{2}$. By making use of equation (\ref{326}) we obtain a square integrable function $H(P)$ which, as a consequence of this fact, satisfies the ONC (\ref{321}):
$$
H(P)=\frac{e^{-iPP'+P'^2/2}}{2\pi^{3/4}}\sum_{\underline l\in\Z^2} f_{\underline l}\int_{-\infty}^{\infty}\,
  \int_{-\infty}^{\infty}e^{-i(xP'+yP+xy/2)}\Phi_{h,\underline l}(x,y)\,dx dy.
$$
After some minor computation and using the integral expression for  $\Phi_{h,\underline l}$, we get:
\be
H(P)=\sum_{\underline l\in\Z^2}f_{\underline l}h(P-al_1)e^{-2iaPl_2}.
\label{410}
\en
In other words, we conclude that, given a seed function $h(P)$, the function $H(P)$ defined as above, with the coefficients $f_{\underline l}$ given in (\ref{49}), satisfies the following ONC
\be
\int_{-\infty}^{\infty}H(P)\overline{H(P-al_1)}e^{-2iaPl_2}\,dP=\int_{-\infty}^{\infty}H(P)\overline{H(P+\tilde X_{\underline l})}e^{iP\tilde Y_{\underline l}}\,dP=\delta_{l_1,0}\delta_{l_2,0}.
\label{411}
\en
We can now use $H(P)$ to find the coefficients of the MRA as in (\ref{322}):
\be
H_n=\frac{1}{\sqrt{a}}\int_{-\infty}^\infty dp e^{-inxa}H(x)=\sqrt{a}\hat H(na),
\label{412}
\en
where $\hat H(p)$ is the Fourier transform of the function $H(x)$.
These coefficients, for what has been discussed in the previous section, automatically satisfy condition (r1): 
\be
\sum_{n\in\Z}H_n\overline{H_{n+2l}}=\delta_{l,0},
\label{413}
\en
simply as a consequence of the (\ref{411}) above. Introducing the Fourier transform of the function $h(x)$, $\hat h(p)$, the integral in (\ref{412}) can be written as:
\be
H_n=\sqrt{a}\sum_{\underline l\in\Z^2}f_{\underline l}\hat h((n+2l_2)a),
\label{414}
\en
which is the expression of the coefficients in terms of the seed function. Making use of the PSF this expression can be  further simplified. In fact, summing over $l_1$, we get 
\be
H_n=\sqrt{a}\sum_{s\in\Z}c_s\hat h((n+2s)a),
\label{415}
\en
where we have defined the new coefficients $c_s$ as follows:
\be
c_s=\frac{1}{2\pi}\int_0^{2\pi}\frac{e^{-ips}\,dp}{\sqrt{S^{(h)}(0,p)}}.
\label{416}
\en

\vspace*{1mm}

\underline{\bf REMARK:-} In the above procedure we have made essentially no requirement on $h(x)$. In particular, we have not assumed that $h$ satisfies the ONC (\ref{321}) from the very beginning, but we have asked $S^{(h)}(0,p)$ to have non zero in $[0,2\pi[$. This is the reason why we had to construct, starting from $h$, a new function $H$ which {\bf does satisfy} the ONC.  It is interesting to remark that, whenever $h$ is already a solution of condition (\ref{321}), $H(x)$ coincides with $h(x)$. Infact, under this assumption, $S^{(h)}_{\underline l}=\delta_{\underline l,\underline 0}$, so that $S^{(h)}(\underline p)=1$. Therefore $f_{\underline l}=\delta_{\underline l,\underline 0}$ and, see (\ref{410}), $H(P)=h(P)$. This will happen, for instance, in Examples 1 and 2 below.

\vspace{3mm}

Before going on considering the other requirements of the relevant sequences, we give the following summation rules, which can be deduced from the definitions above and from the PSF. We have:
\be
\sum_{r_1\in\Z}S_{r_1,r_2}^{(h)}=a\sum_{r_1\in\Z}\hat h(ar_1)\overline{\hat h((r_1-2r_2)a)}, \quad\mbox{ for all fixed } r_2\in\Z;
\label{417}
\en
\be
\sum_{r_2\in\Z}S_{r_1,r_2}^{(h)}=\frac{a}{2}\sum_{r_2\in\Z} h(\frac{ar_2}{2})\overline{h(\frac{a}{2}(r_2-2r_1)}, \quad\mbox{ for all fixed } r_1\in\Z;
\label{418}
\en
\be
\sum_{\underline r\in\Z^2}S_{\underline r}^{(h)}=a\sum_{\underline r\in\Z^2}\hat h(ar_1)\overline{\hat h((r_1-2r_2)a)}=\frac{a}{2}\sum_{\underline r\in\Z^2} h(\frac{ar_2}{2})\overline{h(\frac{a}{2}(r_2-2r_1)}=S^{(h)}(\underline 0);
\label{419}
\en
\be
\sum_{s\in\Z}|c_s|^2=\frac{1}{2\pi}\int_0^{2\pi}\frac{dp}{|S^{(h)}(0,p)|};
\label{420}
\en
\be
\sum_{s\in\Z}c_s=\frac{1}{\sqrt{S^{(h)}(\underline 0)}}.
\label{421}
\en
The proofs of all these equalities are trivial and will not be given here.

% Section 4
\section{The seed function, part two}

In this section we move our attention to the conditions that a seed function $h(x)$ must satisfy in order to produce, via formula (\ref{415}), a set of coefficients $\{H_n\}$ which satisfies conditions (r2)-(r4) of Section II. This will conclude the construction of our relevant sequences. 

\subsection{On the asymptotic behaviour of $H_n$}

We are interested here in finding conditions on $h(x)$ which implies condition (r2). Before considering this problem, it may be interesting to observe that, due to definition (\ref{412}), there exists an easy way to characterize the situation which produces a finite sequence of coefficients $H_n$: using the same notations as in \cite{reed} we say that $H=\{H_n, n\in\Z\}$ belongs to $f$, the set of all the complex sequences with only a finite number of non zero entries, if and only if $\hat H(p)$ is compactly supported. Unfortunately, the analysis of the support of $\hat H(p)$ could be an hard problem, so that this result is of little practical use. More useful is to approach this problem within the framework of convolutions of sequences. We refer to the Appendix for some results on this topic which will be used here. In fact, it is not hard to check that formula (\ref{415}) can be rewritten in terms of convolutions. Defining two sequences related to $\hat h(na)$ as
\be
\hat h^{(even)}_k=\hat h(2ka), \hspace{1cm} \hat h^{(odd)}_k=\hat h((2k+1)a),
\label{51}
\en
which share with $\hat h$ the same asymptotic behaviour, we can write
$H_n=\sqrt{a}\sum_{s\in\Z}c_{s}\hat h((n+2s)a)$ as follows:
\be
\left\{
\begin{array}{ll}
H_{2n}=\sqrt{a}(\overline{c}*\hat h^{(even)})_n,\\
H_{2n+1}=\sqrt{a}(\overline{c}*\hat h^{(odd)})_n,\\
\end{array}
\right.
\label{52}
\en
where we have used that $\overline{c}_s=c_{-s}$ and we have defined $(a*b)_n=\sum_{s\in\Z}a_sb_{n-s}$.

We see from (\ref{52}) that $H_n$ has the same behaviour for large $n$ as $(\overline{c}*\hat h)_n$, where $\hat h_n=\hat h(na)$. In order to get information about the asymptotic behaviour of $H_n$, we therefore have to consider the behaviour of the sequences $\{c_n\}$ and $\{\hat h_n\}$. In particular, the decay features of $\hat h_n$  are given by the explicit expression of the seed function $h(x)$ and of its Fourier transform $\hat h(p)$. The situation is not so simple for the coefficients $c_n$, whose definition (\ref{416}) refers to the function $\sigma(p)=\frac{1}{\sqrt{S^{(h)}(0,p)}}$, and, via ((\ref{321}),(\ref{47})), to the seed function itself. The asymptotic behaviour of the $c_n$ can be deduced using standard techniques in the Fourier series theory: whenever $\sigma(p)$ has $n-1$ continuous derivatives and the n-th derivative has a finite number of discontinuities in $[0,2\pi[$, then the $c_s$ goes like $1/|s|^{n+1}$. Of course, this hypothesis is satisfied whenever $S^{(h)}(0,p)$ is $n$-times differentiable and is strictly positive for $p\in[0,2\pi[$. Instead of finding condition on the seed function for this hypothesis to be satisfied we mention here a class of {\em good} examples which will be discussed in more details in the next section, together with many other examples:

let $k$ be a natural number and let $\hat h_k(p)$ be defined as follows,
\be
\hat h_k(p)=\left\{
\begin{array}{ll}
\frac{1}{\sqrt{(2k+1)a}},\quad p\in[0,(2k+1)a[\\
0 \quad \mbox{ otherwise},\\
\end{array}
\right.
\label{53}
\en
then the related coefficients $H_n^{(k)}$ satisfy condition (r1) for all the values of $k$ and decrease faster than any inverse power of $|n|$, so that they satisfy also condition (r2). This follows from the compact support of $\hat h_k(p)$ and from the  $C^\infty$-nature of the function $\sigma(p)$ generated by $\hat h_k(p)$.

\subsection{About the condition $\sum_{n\in\Z}H_n=\sqrt{2}$}

We want to find here conditions on the seed function $h(x)$ which ensures the validity of condition (r3). Again, we will make use several times of the PSF, which will be assumed to hold.

Under this assumption it is not difficult to prove that

\vspace{2mm}

\noindent
	{\bf Proposition}.--
The set of coefficients (\ref{415}) satisfies condition (r3) if and only if 
\be
\sum_{n\in\Z}\hat h(na)=\sqrt{\frac{2}{a}S^{(h)}(\underline 0)}.
\label{54}
\en

\noindent
	{\bf Proof}

>From the definition (\ref{415}) we see that (r3) is satisfied whenever $\sum_{s,n\in\Z}c_s\hat h((n+2s)a)=\sqrt{\frac{2}{a}}$. Introducing the integer $m=n+2s$ and using equation (\ref{421}), we get equality (\ref{54}). The converse is straightforward. 

\vspace{2mm}

Another result related to this is the following

\vspace{2mm}

\noindent
	{\bf Corollary}.--
Whenever the PSF can be applied, a necessary condition for (r3) to hold is that 
\be
\sum_{n,m\in\Z}\hat h(na)\overline{\left[\hat h(ma)-2\hat h((n-2m)a)\right]}=0
\label{55}
\en

is satisfied. Furthermore, if $\hat h(p)$ has a finite support in $\R$, then the above condition reads
\be
\sum_{n\in\N}(-1)^n\hat h(na)=0
\label{56}
\en

\noindent
	{\bf Proof}

The first statement directly follows from the previous proposition and from equation (\ref{419}).

Formula (\ref{56}) follows from (\ref{55}) and from a direct computation, assuming that $\hat h(p)$ is equal to zero outside a given interval $[-N_1a,N_2a[$, for any $N_1,N_2=0,1,2,3,...$. Under these conditions it is easy to check that   
$$
\sum_{n,m\in\Z}\hat h(na)\overline{\left[\hat h(ma)-2\hat h((n-2m)a)\right]}=-\left|\sum_{n\in\Z}(-1)^n\hat h(na)\right|^2,
$$
so that (\ref{56}) follows.

\subsection{About the condition $H(\omega)\neq 0\quad \forall \omega\in[-\frac{\pi}{2},\frac{\pi}{2}]$}

Let $H(\omega)$ be defined as in (r4),
\be
H(\omega)=\frac{1}{\sqrt{2}}\sum_{n\in\Z}H_ne^{-i\omega n},
\label{57}
\en
with $H_n$ as in  (\ref{415}). Then we can rewrite $H(\omega)$ as follows:
\be
H(\omega)=\sqrt{\frac{a}{2}}{\cal K}(2\omega){\cal H}(-\omega), \,\mbox{ where } \,\,{\cal K}(\omega)=\sum_{s\in\Z}c_se^{i\omega s}, \,\,{\cal H}(\omega)=\sum_{s\in\Z}\hat h(sa)^{i\omega s}.
\label{58}
\en
Due to the equality $c_{-s}=\overline{c}_s$ we can check that ${\cal K}$ is a real function. Moreover, we can also check that ${\cal K}(2\omega)\neq 0$ for all $\omega\in[-\frac{\pi}{2},\frac{\pi}{2}]$ or, equivalently, that ${\cal K}(\nu-\pi)\neq 0$ for all $\nu\in[0,2\pi]$. The proof of this statement follows again from the PSF. In particular we can check that
\be
{\cal K}(\nu-\pi)=\left\{
\begin{array}{ll}
\frac{1}{\sqrt{S^{(h)}(0,\nu+\pi)}}, \mbox{ if } 0\leq\nu<\pi\\
\frac{1}{\sqrt{S^{(h)}(0,\nu-\pi)}}, \mbox{ if } \pi\leq\nu\leq2\pi,\\
\end{array}
\right.
\label{59}
\en
and for this reason $H(\omega)$ is different from $0$ in $[-\frac{\pi}{2},\frac{\pi}{2}]$ if and only if 
${\cal H}(\omega)\neq 0$ in $[-\frac{\pi}{2},\frac{\pi}{2}]$, condition  which is easier to be verified since it is directly linked to the seed function $\hat h(p)$. In the next section we will discuss examples of seed functions satisfying this condition.

\underline{\bf Remarks:}-- (1) One can think that analogous results could be obtained in a completely different (and, maybe, more natural) way, that is by starting from a given {\em seed sequence} $\{h_n,\, n\in\Z\}$, normalized in $l^2(\Z)$, and by defining a new sequence $H_n=\sum_{s\in\Z}c_sh_{n+s}$. The problem should be now finding conditions on $c_s$ such that properties (r1)-(r4) are satisfied. It is not very hard to check that, even if this approach does not seem to be very different from what we have done, it is quite difficult to obtain reasonable conditions on $c_s$: what is missing, from our point of view, is the possibility of mapping the problem into a complete different settings, in which the requirement $\sum_{n\in\Z}H_n\overline{H}_{n+2l}=\delta_{l,0}$ can be considered simply as an orthonormality requirement between wave-functions in a certain subspace of $\Lc^2(\R^2)$.

(2)  It may be useful to remark also that the generic use of the sentence {\em whenever the PSF holds} is related to the fact that several inequivalent hypotheses could be checked in order to ensure the validity of the PSF. For instance, multiplying  formula (\ref{324}) for a function $\varphi(x)$ and integrating over $\R$, we know that the equality holds for instance (1) if $\varphi$ belongs to $\Sc$ or (2) if  $\varphi$ belongs to $\Lc^1(\R)$ and is continuous and with bounded variation or (3) if $\varphi$ is continuous and if $\sup_{x\in\R}(|\varphi(x)|+|\hat \varphi(x)|)(1+|x|)^{1+\epsilon}<\infty$. Moreover, we will find in the next section other situations in which none of these conditions are satisfied but, nevertheless, the validity of the PSF can be explicitly proved. In conclusion, we find that the most economical way to handle with the PSF is simply to check its validity whenever is needed.

%\section5
\section{Examples}

This section is devoted to the analysis of several applications of the construction outlined in Sections III and IV.

\underline{\bf Example 1.}

Let us consider the following function, defined in the momentum space:
\be
\hat h(p)=\left\{
\begin{array}{ll}
\frac{1}{\sqrt{a}},\quad p\in[0,a[\\
0 \quad \mbox{ otherwise.}\\
\end{array}
\right.
\label{61}
\en
This is a normalized function in $\Lc^2(\R)$, and the coefficients $S^{(h)}_{\underline l}$, defined as in (\ref{321}), are all zero but when $l_1=l_2=0$: $S^{(h)}_{\underline l}=\delta_{\underline l,\underline 0}$. Therefore $S^{(h)}(\underline p)=1$ and, as a consequence of (\ref{416}), $c_s=\delta_{s,0}$. Therefore $H_n=\sqrt{a}\hat h(na)=\sqrt{a}\delta_{n,0}$, which clearly satisfies (r1), (r3) and (r4) but does not satisfy condition (r2). Furthermore, it is easy to check that all the sum rules given in Section III are satisfied.
For instance, it is straightforward to check explicitly equation (\ref{421}). This shows that the PSF can be applied also for a function $\sigma(p)=1$, which does not fit any of the hypotheses given before.

\underline{\bf Example 2.}

Let us consider the following function, defined again in the momentum space:
\be
\hat h(p)=\left\{
\begin{array}{ll}
\frac{1}{\sqrt{2a}},\quad p\in[0,2a[\\
0 \quad \mbox{ otherwise.}\\
\end{array}
\right.
\label{62}
\en
As before we find $S^{(h)}_{\underline l}=\delta_{\underline l,\underline 0}$, $S^{(h)}(\underline p)=1$ and $c_s=\delta_{s,0}$. Therefore, $H_n=\sqrt{a}\hat h(na)=\frac{1}{\sqrt{2}}(\delta_{n,0}+\delta_{n,1})$. We have obtained therefore the coefficients of the Haar MRA: all the properties (r1)-(r4) are obviously satisfied, as well as all the sum rules given before. 

We want to remark that in both these examples the ONC (\ref{321}) was already satisfied by the seed function itself, and for this reason it is not a surprise that the new function $H$ in (\ref{410}) coincides with $h$.

\underline{\bf Example 3.}

Let us consider:
\be
 h(x)=\left\{
\begin{array}{ll}
\frac{1}{\sqrt{da}},\quad p\in[0,da[\\
0 \quad \mbox{ otherwise,}\\
\end{array}
\right.
\label{63}
\en
where $d=1,2,3,..$. This time the seed function has compact support in the position space, so that $\hat h(p)$ decayes  rather slowly.  $S^{(h)}(0,p)$ is, in general, different from 1 but is independent of $p$, so that $c_s$ is again proportional to $\delta_{s,0}$. Moreover an explicit computation shows that $\hat h(na)$ is different from zero only if $n=0$, so that $H_n$ turns out to be non zero only if $n=0$. Therefore, even if the seed function is quite different from that of Example 1, the resulting coefficients essentially coincide with those obtained there. The sum rules again are verified.

\underline{\bf Example 4.}

Let us define now:
\be
\hat h(p)=\left\{
\begin{array}{ll}
\frac{1}{\sqrt{3a}},\quad p\in[0,3a[\\
0 \quad \mbox{ otherwise.}\\
\end{array}
\right.
\label{64}
\en
We get easily $S^{(h)}_{\underline r}=\delta_{r_1,0}\left[\delta_{r_2,0}+\frac{1}{3}(\delta_{r_2,1}+\delta_{r_2,-1})\right]$ which implies that $S^{(h)}(\underline p)=1+\frac{2}{3}\cos(p_2)$. We see that $S^{(h)}(0,p)$ is always positive in $[0,2\pi]$ and infinitely differentiable. We can deduce, therefore, that the $c_s$' decay faster than any inverse power of $|s|$. Since $\hat h(p)$ is different from zero only in the finite set $[0,3a[$ we can use the result of the Proposition given in the Appendix, statement 1), to conclude that the sequence $H_n$ in (\ref{415}) satisfies conditions (r1) and (r2). However,  since (\ref{54}) is not verified, we do not expect condition (r3) to hold. All the sum rules can be explicitly checked.

\underline{\bf Example 5.}

Let $h(x)=\frac{1}{\pi^{1/4}}e^{-x^2/2}$. Its Fourier transform is $\hat h(p)=\frac{1}{\pi^{1/4}}e^{-p^2/2}$. Using formula (\ref{321}) we find $S^{(h)}_{\underline r}=e^{-\frac{\pi}{2}(r_1^2+4r_2^2)}$, which implies that $S^{(h)}(0,p)=\sum_{r_1\in\Z}e^{-\frac{\pi}{2}r_1^2}\sum_{r_2\in\Z}e^{-2\pi r_2^2}e^{ipr_2}$. The sum in $r_1$ can be performed numerically and it gives $\sum_{r_1\in\Z}e^{-\frac{\pi}{2}r_1^2}= 1.4195$. Using now the usual techniques outlined above and in the Appendix we can easily deduce that, not only condition (r1) but also conditions (r2) and (r4) are automatically satisfied, the reason being the very fast decay properties of both $c_s$ and $\hat h$. However, condition (r3) is not verified since equality (\ref{54}) does not hold. On the contrary, all the sum rules deduced in Section III are verified.

Let us work out this example in more details. Since the explicit computation of $S^{(h)}(0,p)$ is difficult, we  consider here a perturbative computation. We will show  that already a very crude approximation gives interesting results, and that a slightly better approximation makes the result almost exact. The main difficulty consists in the computation of $c_s$ in (\ref{416}). Using the expansion $\frac{1}{\sqrt{1+x}}=1-\frac{1}{2}x+\frac{3}{8}x^2+..$,  and observing that $\sum_{r_2=1}^\infty e^{-2\pi r_2^2}\simeq 0.00186$, we can proced as follows:
\beano
&&\frac{1}{\sqrt{S^{(h)}(0,p)}}= \frac{1}{\sqrt{1.4195}}\frac{1}{\sqrt{1+2\sum_{r_2=1}^\infty e^{-2\pi r_2^2}\cos(pr_2)}}\simeq \\&&\simeq\frac{1}{\sqrt{1.4195}}\left(1-\sum_{r_2=1}^\infty e^{-2\pi r_2^2}\cos(pr_2)\right) \simeq \frac{1}{\sqrt{1.4195}}
\enano
considering the crudest approximation (the rest is only 2/1000 of the main contribution!). In this way we get $c_s\simeq \frac{\delta_{s,0}}{\sqrt{1.4195}}$, and therefore $H_n\simeq\frac{2^{1/4}}{\sqrt{1.4195}}e^{-\pi n^2}$. It is clear that both (r2) and (r4) are satisfied. As for the (r1), a numerical computation shows that $\sum_{n\in\Z}H_n^2\simeq 0.999992$,  $\sum_{n\in\Z}H_nH_{n\pm 2}\simeq 0.00186$, and $\sum_{n\in\Z}H_nH_{n\pm 2l}$ is even smaller for $|l|$ larger than 1. We see that this is already a good approximation of (\ref{323}). Better results can be obtained simply considering the next contribution in the previous expansion, which means considering also the term with $r_2=1$ in the sum above. In this case we get $c_s\simeq \frac{1}{\sqrt{1.4195}} (\delta_{s,0}-\frac{1}{2}(\delta_{s,1}+\delta_{s,-1}))$, and 
$$
H_n\simeq
\frac{2^{1/4}}{\sqrt{1.4195}}\left[e^{-\pi n^2}-\frac{1}{2}e^{-2\pi}(e^{-\pi (n+2)^2}+e^{-\pi (n-2)^2})\right].
$$
We find now that $\sum_{n\in\Z}H_n^2\simeq 0.999992$, while $\sum_{n\in\Z}H_nH_{n\pm 2}\simeq 10^{-8}$ which is much smaller than before.
As for (r3), a numerical computation gives $\sum_{n\in\Z}H_n\simeq 1.0844\neq \sqrt{2}$, as expected. Again, all the sum rules are satisfied.

\underline{\bf Example 6.}

This example generalizes Example 2 above, in the sense that we still require $\hat h(p)$ to be zero outside $[0,2a[$ but we do not fix the analitic expression of $\hat h$ inside $[0,2a[$. Without going in all the details we just want to remark that also now $c_s$ is proportional to $\delta_{s,0}$, so that $H_n$ is proportional to $\hat h(na)$. More in detail we find 
$$
H_n=\frac{1}{\sqrt{|\hat h(0)|^2+|\hat h(a)|^2}}\left(\hat h(0)\delta_{n,0}+\hat h(a)\delta_{n,1}\right).
$$
It is clear that conditions (r1), (r2) and (r4) are automatically satisfied, while (r3) holds whenever $\hat h(p)$ is such that  $\frac{\hat h(0)+\hat h(a)}{\sqrt{|\hat h(0)|^2+|\hat h(a)|^2}}=\sqrt{2}$.

\underline{\bf Example 7.}

This examples can be considered as a generalization of Examples 1 and 4 and was already mentioned in the previous section. Let $k$ be a fixed natural: $k=0,1,2,..$, and let
\be
\hat h_k(p)=\left\{
\begin{array}{ll}
\frac{1}{\sqrt{(2k+1)a}},\quad p\in[0,(2k+1)a[\\
0 \quad \mbox{ otherwise.}\\
\end{array}
\right.
\label{65}
\en
Obviously, $k=0$ returns Example 1, while $k=1$ gives Example 4. Computing the integral in (\ref{321}) we find 
$$
S^{(h_k)}(0,p)=1+(1-\delta_{k,0})\frac{2}{2k+1}\sum_{j=0}^{k-1}(2j+1)\cos(p(k-j)),
$$
which turns out to be strictly positive for all values of $k$. This claim was analitically and numerically checked for many values of $k$. For $k$ increasing it is possible to see that the function $S^{(h_k)}(0,p)$ approaches more and more zero, but, at least for $k\leq 100$, it is always strictly positive. We guess that this same positivity also holds for $k$ bigger than 100, but an analytical control is quite difficult in this case and it is not very relevant here. Incidentally, this is the reason why the seed function $\hat h_k(p)$ is defined on, say, odd intervals. For even ones, in fact, ($p\in[0,2ka[$), it is easy to check that $S^{(h_k)}(0,p)$ has a zero inside $[0,2\pi[$, and the integral defining $c_s$ diverges.

It is now clear that, for any fixed $k$, the function $\frac{1}{S^{(h_k)}(0,p)}$ is in $C^\infty$, so that $c_s$ decays faster than any inverse power of $|s|$. Now, since $\hat h(p)$ is different from zero only in a finite interval, it is clear also that for the asymptotic behaviour of the  coefficients $H_n=\sqrt{a}\sum_{s\in\Z}c_s\hat h((n+2s)a)$ we can apply the Proposition given in the Appendix, statement 1), so that we conclude that $H_n\in\Scmin$, where $\Scmin$ is defined in the Appendix. Condition (r3) does not hold since equation (\ref{54}) is not verified.

\underline{\bf Example 8.}

Let us fix $l\in\N$ and define 
\be
h_l(x)=\left\{
\begin{array}{ll}
\sqrt{\frac{2}{la}},\quad x\in[0,\frac{la}{2}[\\
0 \quad \mbox{ otherwise.}\\
\end{array}
\right.
\label{66}
\en
This class of seed functions is interesting because it produces, after the usual procedure, a set of coefficients $c_s$ which are always zero but if $s=0$. Therefore we obtain:
$$
H_n=\sqrt{\frac{a}{S^{(h_l)}(\underline 0)}}\hat h_l(na). 
$$
Whenever $l$ is even the situation is not very interesting, since we get $H_n\propto\delta_{n,0}$. On the contrary, if $l$ is odd, $l=2k+1$, we find that
$$
\hat h_{2k+1}(na)=\left\{
\begin{array}{ll}
\sqrt{\frac{2k+1}{2a}},\hspace{2cm}&n=0\\
0  &n=\pm 2,\pm 4,\pm 6,..\\
\sqrt{\frac{2}{ina\sqrt{\pi a(2k+1)}}}, &n=\pm 1,\pm 3,\pm 5,..\\
\end{array}
\right.
$$
We see therefore that, even if (r1) is satisfied, (r2) is not. Also (r3) does not hold since equation (\ref{54}) is not verified.

\underline{\bf Example 9.}

As a final example here we consider the following seed function $\hat h(p)=\frac{2}{a(1+p^2)}$, which produces the following coefficients $S^{(h)}_{\underline r}=\frac{e^{-|r_1|a}}{1+2\pi r_2^2}$ and the following function $S^{(h)}(\underline p)$:
$$
 S^{(h)}(\underline p)=\frac{1+e^{-a}}{1-e^{-a}}\varphi(p_2), \quad \mbox{ with }\quad \varphi(p_2)=\sum_{r_2\in\Z}\frac{e^{ip_2r_2}}{1+2\pi r_2^2}.
$$
It is an easy estimate to check that $\varphi(p_2)\neq 0$ in $[0,2\pi[$. However, we cannot use the same arguments as for the Example 5 to conclude that $\varphi(p_2)$ belongs to $C^\infty$, the reason being that the Fourier coefficients $\frac{1}{1+2\pi r_2^2}$ of $\varphi$ do not decay very fast. For this reason is not difficult to understand that condition (r2) is not satisfied whereas conditions (r1) and (r3) hold. In particular this last condition can be controlled by checking directly equation (\ref{54}).

\vspace{3mm}

Let us now  go back to equation (\ref{48}), where the phase $\varphi(\underline p)$ was chosen to be equal to zero. We want to show here that this is really a very special choice. Infact, the following  two simple examples point out that a different choice of $\varphi(\underline p)$ produces coefficients $H_n$ which can be significantly different from the ones we get if $\varphi(\underline p)=0$. 

First we remark that the expression for $c_s$ must be a little bit modified. Instead of (\ref{416}) we have
\be
c_s=\frac{1}{2\pi}\int_0^{2\pi}\frac{e^{-ips+i\varphi(0,p)}\,dp}{\sqrt{S^{(h)}(0,p)}}.
\label{67}
\en
A first application of this formula consists in choosing $\varphi(0,p)=pK_0$, $K_0$ being a fixed integer. If we consider, for instance, Example 2 above, we see that the only difference, in this case, is that, instead of having $c_s=\delta_{s,0}$, we find $c_s=\delta_{s,K_0}$, so that $H_n=\frac{1}{\sqrt{2}}(\delta_{n,K_0}+\delta_{n,K_0+1})$. More interesting is the situation if $\varphi$ is not linear. Let us consider here $\varphi(0,p)=\gamma p^2$, $\gamma\in\R$. Restricting again to Example 2, for which $S^{(h)}(\underline 0)=1$, we can still compute analitically the coefficients $c_s$ which turn out to be
$$
c_s=\frac{-1}{4\pi}\sqrt{\frac{\pi}{-i\gamma}}e^{-\frac{is^2}{4\gamma}}\left(\Phi\left(\frac{i(4\pi\gamma-s)}{2\sqrt{-i\gamma}}\right)+\Phi\left(\frac{is}{2\sqrt{-i\gamma}}\right)\right),
$$
where $\Phi$ is the erf function, \cite{grad}. Using its well known asymptotic behaviour, we find that $c_s$ decays as $|s|^{-1}$, that is a very slow behaviour when compared with that obtained for $\varphi=0$.

\section{Conclusions}

We have shown how to use the relation  between the FQHE and the MRA recently established by the author in order to construct a set of coefficients which produce a MRA of $\Lc^2(\R)$. The examples given show that, while is essentially automatic to obtain a sequence satisying condition (r1), more care must be used to find a seed function which produce a relevant sequence. Conditions on the seed function for the set $\{H_n, n\in\Z\}$ to be relevant are discussed.

%\newpage
\section*{Acknowledgements}

This work has been financially supported in part by M.U.R.S.T.,
within the  project {\em Problemi Matematici Non Lineari di
Propagazione e Stabilit\`a nei Modelli del Continuo}, coordinated
by Prof. T. Ruggeri.

\vspace{.4cm}

 \appendix

\renewcommand{\theequation}{\Alph{section}.\arabic{equation}}

 \section{\hspace{-.7cm}ppendix :  Convolutions of sequences}

In this Appendix we prove some results concerning the asymptotic behaviour of convolutions in view of applications. We wish to stress that these results are given here since, though being quite reasonable, were not found by the author in the existing literature.

We use here the same notation as in \cite{reed}: $f$, $s$ and $l_p$ are well known spaces of sequences, the first containing all the {\em finite} sequences, that is, those sequences which are zero outside of a finite set of indexes. The other sets are defined as follows:
\be
s=\{a: \lim_{|n|,\infty}|n|^pa_n=0, \quad\forall p\in \N\},\quad
l_p=\{a: \|a\|_p=(\sum_{p\in\Z}|a_n|^p)^{1/p}<\infty\}.
\label{a1}
\en
Given two sequences $a,b$ we define a third sequence $c=a*b$ as $c_n=\sum_{s\in\Z}a_sb_{n-s}=\sum_{s\in\Z}a_{n-s}b_{s}$. We have the following

\noindent
	{\bf Proposition}.--
Let $a,b$ and $c$ be as above. Then the following statements hold:

1) if $a\in f$ then the asymptotic behaviour of $c$ is the same of that of $b$;

2) if $a\in l_1$ and $b\in l_p$ then $c\in l_p$, for all $1\leq p<\infty$;

3) if $a,b\in s$ then $c\in s$.

\noindent
	{\bf Proof}

1) This is clear because $a_n=0$ but for a finite number of indexes $n$. Of course the same result can be related simply by exchanging the roles of $a$ and $b$.

2) The proof of this statement follows from well known properties of the convolutions of functions. We start defining two functions, defined in $\R$, as follows:
$$
a(x)=|a_s|, \quad x\in[s,s+1[, \hspace{2cm} b(x)=|b_s|, \quad x\in[s,s+1[, \quad s\in\Z.
$$
It is clear that $a(x)\in\Lc^1(\R)$, while $b(x)\in\Lc^p(\R)$. Then it is well known that $a*b\in\Lc^p(\R)$, where $(a*b)(x)=\int_{\R}a(y)b(x-y)dy$. In order to conclude that $c\in l_p$ we consider that
$$
c(x)=\int_{\R}a(y)b(x-y)dy=\sum_{s\in\Z}\int_s^{s+1}a(y)b(x-y)dy=\sum_{s\in\Z}|a_s|\int_s^{s+1}b(x-y)dy.
$$
Using now the definition of $b(x)$ it is easy to check that, for all integer $l$ and for $0\leq\alpha<1$, we have 
\be
c(l+\alpha)=\sum_{s\in\Z}|a_s|\left((1-\alpha)|b_{l-s-1}|+\alpha|b_{l-s}|\right)=(1-\alpha)d_{l-1}+\alpha d_l,
\label{a2}
\en
where we have defined $d_l=\sum_{s\in\Z}|a_sb_{l-s}|\geq 0$, for all $l\in\Z$. The conclusion now follows from the fact that $c(x)$ belongs to $\Lc^p(\R)$ and from the inequality $(\gamma_1+\gamma_2+...\gamma_n)^p\geq \gamma_1^p+\gamma_2^p+...\gamma_n^p$, which holds whenever $\gamma_j\geq 0$ and for all $p\geq 1$. In fact we have:
$$
\begin{array}{ll}
\infty>\int_{\R}|c(x)|^pdx=\sum_{l\in\Z}\int_l^{l+1} |c(x)|^pdx=\sum_{l\in\Z}\int_0^{1} |c(l+x)|^pdx=\\
=\sum_{l\in\Z}\int_0^{1} ((1-\alpha)d_{l-1}+\alpha d_l)^pd\alpha\geq \frac{2}{p+1}\sum_{l\in\Z}d_l^p\geq \frac{2}{p+1}\sum_{l\in\Z}c_l^p,
\end{array}
$$
which proves that $c\in l_p$.

3) From the definition $c_n=\sum_{s\in\Z}a_sb_{n-s}$ we get easily the following equality between functions: $C(p)=A(p)B(p)$, where $A(p)=\sum_{s\in\Z}a_se^{isp}$, $B(p)=\sum_{s\in\Z}b_se^{isp}$
 and $C(p)=\sum_{s\in\Z}c_se^{isp}$. The coefficients $c_l$ can now be found simply by
\be
c_l=\frac{1}{2\pi}\int_0^{2\pi}C(p)e^{-ipl}dp=\frac{1}{2\pi}\int_0^{2\pi}A(p)B(p)e^{-ipl}dp,
\label{a3}
\en
which is the starting point of our asymptotic analysis. In fact, due to the fact that $a,b\in s$, the functions $A(p)$ and $B(p)$ belongs to $C^\infty$, and so their product does. This implies, using well known fact about the Fourier series, that the coefficients $c_l$ in (\ref{a3}) decay faster than every inverse power of $|l|$, so that $c\in s$.

\hfill

\underline{\bf Remark:}-- It is clear that statement 2) is not enough to ensure validity of (r2), which is satisfied, on the contrary, if $a$ and $b$ are both in $s$ or if, e.g., $a$ is in $f$ and $b$ decays like $1/n^2$.

\end{document}